\documentclass[english,twocolumn,floatfix,prl,aps,showpacs,bibliography]{revtex4-1}
\usepackage[T1]{fontenc}
\usepackage[latin9]{inputenc}
\usepackage{babel}
\usepackage{amsbsy}
\usepackage{amssymb}
\usepackage{graphicx}
\usepackage[unicode=true,pdfusetitle,
 bookmarks=true,bookmarksnumbered=false,bookmarksopen=false,
 breaklinks=false,pdfborder={0 0 1},backref=false,colorlinks=false]
 {hyperref}
 
\makeatletter

\usepackage{graphicx,color,curves,epic}

\makeatother

\begin{document}

\newcommand\ket[1]{\left|#1\right\rangle }
\newcommand\bra[1]{\left\langle #1\right|}
\newcommand\braket[2]{\left\langle #1\middle|#2\right\rangle }
\newcommand\ketbra[2]{\left|#1\vphantom{#2}\right\rangle \left\langle \vphantom{#1}#2\right|}
\newcommand\braOket[3]{\left\langle #1\middle|#2\middle|#3\right\rangle }

\newcommand{\FIXME}[1]{({\bf FIXME: #1})}
\newcommand{\ie}{\emph{i.e.}}
\newcommand{\eg}{\emph{e.g.}}
\renewcommand{\i}{\imath}

\title{The existence of robust edge currents in Sierpinsky Fractals}
\author{Mikael Fremling$^{1}$, Michal van Hooft$^{1}$, Cristiane Morais Smith$^{1}$, Lars Fritz$^{1}$}

\affiliation{$^{1}$Institute for Theoretical Physics, Center for Extreme Matter
and Emergent Phenomena, Utrecht University, Princetonplein 5, 3584
CC Utrecht, the Netherlands}

\begin{abstract}
We investigate the Hall conductivity in a Sierpinski carpet,
a fractal of Hausdorff dimension $d_f=\ln(8)/\ln(3) \approx 1.893$,
subject to a perpendicular magnetic field.
We compute the Hall conductivity using linear response and the recursive Green function method.
Our main finding is that edge modes, corresponding to a maximum Hall conductivity of at least $\sigma_{xy}=\pm \frac{e^2}{h}$,
seems to be generically present for arbitrary finite field strength,
no mater how one approaches the thermodynamic limit of the fractal.
We discuss a simple counting rule to determine the maximal number of edge modes  in terms of paths through the system with a fixed width.
This quantized edge conductance,
as in the case of the conventional Hofstadter problem,
is stable with respect to disorder and thus a robust feature of the system.
\end{abstract}
\maketitle

\begin{figure}[b]
  \begin{centering}
    \includegraphics[width=0.95\columnwidth]{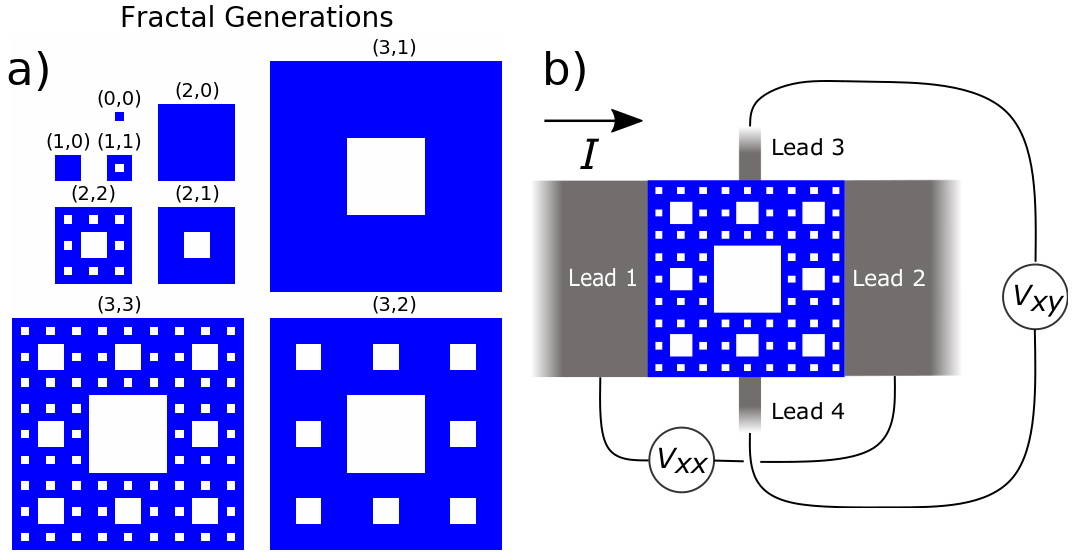}
\par\end{centering}
\caption{(a) Iterative construction of the fractals. We have $g \geq f \geq 0$; (b) schematic of the four terminal Hall bar setup.
  It allows to measure the resistance between lead 1 and lead 2, $\rho_{xx}=V_{xx}/I$,
and the Hall resistance $\rho_{xy}=V_{xy}/I$ between lead 3 and lead 4.\label{fig:Fracs_and_Transport_Setup}}
\end{figure}

\begin{figure*}[!t]
  \begin{centering}
    \includegraphics[width=1.0\linewidth]{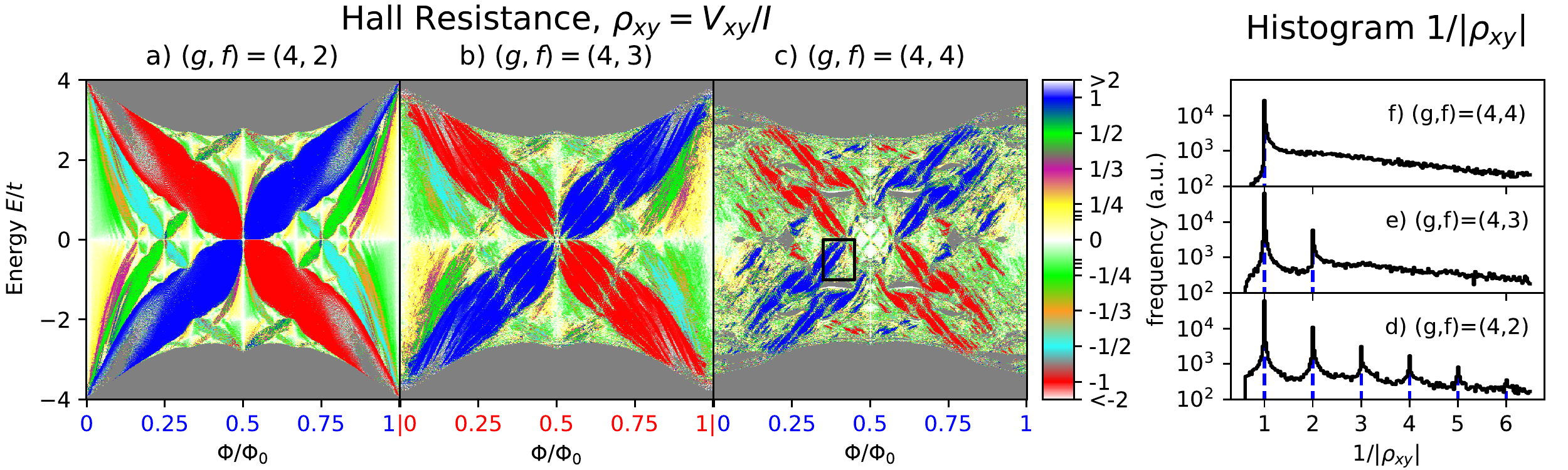}
    \par\end{centering}
\caption{Hall butterfly: The main panels show the Hall resistance $\rho_{xy}=\frac{V_{xy}}I$ measured between lead 3 and lead 4, if a current is driven from lead 1 to lead 2,
  see Fig.~\ref{fig:Fracs_and_Transport_Setup}.
  The color-scheme is truncated at $-2<\rho_{xy}<2$,
  and grey areas have total transmission coefficient lower than $0.1$.
  The right panels show the histogram of $\frac{1}{|\rho_{xy}|}$ over the whole parameter space.
  For $(4,2)$ we can see several peaks of $1/|\rho_{xy}|$ corresponding to multiple edge modes.
  For $(4,3)$ many of these peaks are quenched, and only $1/|\rho_{xy}|=1 ,2$ and possibly $1/|\rho_{xy}|=3$ have peaks.
  Finally, at $(4,4)$, only the $|\rho_{xy}|=1$ peak is preserved.
  The black square shows a region that is studied in further detail in Fig.~\ref{fig:Hall_Butterfly_Zoom_in}.
  \label{fig:Hall_Butterfly}}
\end{figure*}

Topological insulators and their properties, such as exotic surface states, have been at the forefront of condensed matter research during the last decade~\cite{bernevig2013topological}.
Under the most generic circumstances they have been classified in 'the ten-fold way'~\cite{ryu2010topological,qi2011topological} or,
the 'periodic table of topological insulators'~\cite{kitaev2009periodic}.
The key insight is that given the non-spatial symmetries, such as chirality, time-reversal symmetry, and parity,
one can deduce the possible existence or non-existence of states with non-trivial bulk topological order. 

Numerous extensions of this classification scheme have been discussed: imposing lattice symmetries leading to crystalline topological insulators~\cite{fu2011topological},
dissipation in non-Hermitian Hamiltonians~\cite{diehl2011topology},
driven non-equilibrium systems~\cite{lindner2011floquet},
or the so called higher order topological insulators~\cite{benalcazar2017quantized,schindler2018higher}. 

We explore an alternative path which is to modify dimension.
We are especially interested in knowing to which extent one can employ the classifications of the integer dimensions above and below,
when considering a fractional dimension.  
It has been known for a long time that fractal structures possess non-integer Hausdorff dimension $d_f$~\cite{mcmullen1984hausdorff}. Recently,
lattices with fractal structure have been manufactured in the laboratory in a number of ways.
This includes the use of molecular assembly~\cite{Shang2015,Tait2015,Nieckarz2016,Jiang2017,Sun2015,Zhang2015,Zhang2016}, templating, and co-assembly methods~\cite{Li2017a,Li2017b}.
Furthermore, electronic fractal lattices were created by scanning tunneling microscope techniques~\cite{Kempkes2019}. Theoretically,
this sparked interest in electronic~\cite{vanVeen2016} and optical conductivity~\cite{vanVeen2017} as well as plasmon dispersion relations~\cite{Westerhout2018}.
Recent works explored Anderson localization and critical points on fractals~\cite{Sticlet2016} as well as level statistics~\cite{Iliasov2019}.
The prospect of topological order in fractals was investigated in Refs.~\cite{Aoki1990,Aoki1992} and recently revived in Ref.~\onlinecite{Brzezinska2018} and we compare our results to this latter work.

{\it Structure of paper and main results:} We concentrate on the Sierpinski carpet (SC) with a Hausdorff dimension $d_f=\ln(8)/\ln(3)\approx 1.893$~\cite{mcmullen1984hausdorff}.
We start from a two dimensional Chern insulator,
the Hofstadter model, and then convert it into the SC by diluting the lattice.
Subsequently, we employ the recursive Green function method to investigate the Hall resistance and relate it to the number of existing edge modes.
The findings are compared to Chern number calculations.
Our principal finding is that, regardless of the protocol for constructing the fractal,
one generically finds Hall conductances of at least $\sigma_{xy}=\pm e^2/h$ surviving the extrapolation to the thermodynamic limit.
We back this up using several extrapolations schemes.
Furthermore, we show that as in the case of more conventional quantum Hall systems,
this mode is stable upon introducing disorder.

{\it Constructing the fractals:} We distinguish fractal generation $g$,
related to the linear size of the system as $L=3^{g}$,
and fractal depth $f$, counting the number of times the cutting procedure has been iterated,
see Fig.~\ref{fig:Fracs_and_Transport_Setup}.
Both numbers are summarized in a fractal index $\mathcal{F}=\left(g,f\right)$.
The number of sites in a fractal $\mathcal{F}$ is given by $N=L^{2}\left(8/9\right)^{f}=9^{g-f}8^{f}$. The relevant linear sizes,
Hilbert space sizes, and volume fractions can be found in Table~\ref{tab:Table-of-Sizes}.

\begin{table}
\begin{centering}
\begin{tabular}{|c|r|r|r|r|r|r|r|}
\hline 
$g$ & 0 & 1 & 2 & 3 & 4 & 5 & 6\tabularnewline
\hline 
\hline 
$L=3^{g}$ & 1 & 3 & 9 & 27 & 81 & 243 & 729\tabularnewline
\hline 
Dim$\left(H\right)=8^{g}$ & 1 & 8 & 64 & 512 & 4 096 & 32 768 & 262 144\tabularnewline
\hline 
Vol\%$=\left(8/9\right)^{g}$ & 1 & 89\% & 79\% & 70\% & 62\% & 55\% & 49\%\tabularnewline
\hline 
\end{tabular}
\par\end{centering}
\caption{Table of linear system sizes, Hilbert space sizes and volume-fraction of the Sierpinski lattice $\left(g,g\right)$.\label{tab:Table-of-Sizes}}
\end{table}

{\it Model and setup:} We consider a tight-binding Hamiltonian for spinless fermions 
\begin{equation}
  H_{\mathcal{L}}=-t\sum_{\left\langle i,j\right\rangle \in\mathcal{L}}
  \left( a_i^{\dagger} a_j e^{\i A_{ij}}+\mathrm{h.c.} \right),
  \label{eq:Frac_ham}
\end{equation}
subject to a perpendicular magnetic field implemented via $A_{ij}=\int_{{\bf r}_i}^{{\bf r}_j} \vec A\cdot d\vec l$ (we choose the Landau gauge $\vec A=B(y,0)$).
Here $a_i^{\dagger}$ creates a fermion on lattice site $i$ and $\mathcal{L}$ is the set of nearest neighbors with support on the lattice.

The starting point of our construction is a square lattice with lattice spacing $a$.
We parametrize the strength of the magnetic field as $B=2\pi\Phi/a^2$, where $\Phi$ is the magnetic flux piercing every plaquette ($\Phi_0=h/e$ is the elementary flux).

{\it{Transport calculation:}} Instead of studying the open boundary spectrum or topological indices,
we directly probe transport properties.
We compute transport through a Hall bar, see Fig.~\ref{fig:Fracs_and_Transport_Setup}b,
using linear response and the recursive Green function method which gives us access to larger system sizes~\cite{datta1997electronic}.
In a two-terminal setup extended bulk states also contribute to the longitudinal transport, making it difficult to isolate boundary modes.
Consequently, we implement a four-terminal setup giving direct access to the boundaries.
We calculate two transport quantities,
$\rho_{xy}=V_{xy}/I$ and $\rho_{xx}=V_{xx}/I$ ($\rho_{xx}$ is measured between lead 1 and lead 2 and not along the boundary, like in a six terminal setup).
The metallic leads are semi-infinite and described by the Hamiltonian $-t\sum_{\left\langle i,j\right\rangle}a_{i}^{\dagger}a_{j}+\mathrm{h.c}$ describing nearest-neighbor hopping on a square lattice.
The top and bottom leads are 5-10 lattice sites wide.
We use a recursive scheme to calculate the Green function, allowing to target larger $g$ and $f$.

{\it The thermodynamic limit:} We are interested in transport through the fractal in the thermodynamic limit.
There are many different ways to approach the thermodynamic limit.
For instance, all the sequences $\mathcal{F}=\lim_{g\to\infty}(g,g-\delta)$, with $\delta$ finite,
have the same Hausdorff dimension,
and thus in the thermodynamic limit all constitute lattice realizations of the Sierpinski carpet, albeit with different UV-cutoffs.

In the following, we concentrate on two extreme cases:
(i) starting from a large $g$ and following the sequence
$\mathcal{F}=\left(g,0\right)\to\left(g,1\right)\to\left(g,2\right)\to\cdots\to\left(g,g\right)$;
(ii) starting from a small $\mathcal{F}=\left(0,0\right)$ fractal and recursively building larger fractals,
\ie~following the sequence $\mathcal{F}=\left(0,0\right)\to\left(1,1\right)\to\left(2,2\right)\to\cdots\to\left(g,g\right)$.
We found heuristically that sequence (ii) is the one with the lowest number of stable edge states.
We show that irrespective of the UV regularization we choose,
at least one edge mode survives.

\begin{figure}
  \begin{centering}
    \includegraphics[width=1.0\columnwidth]{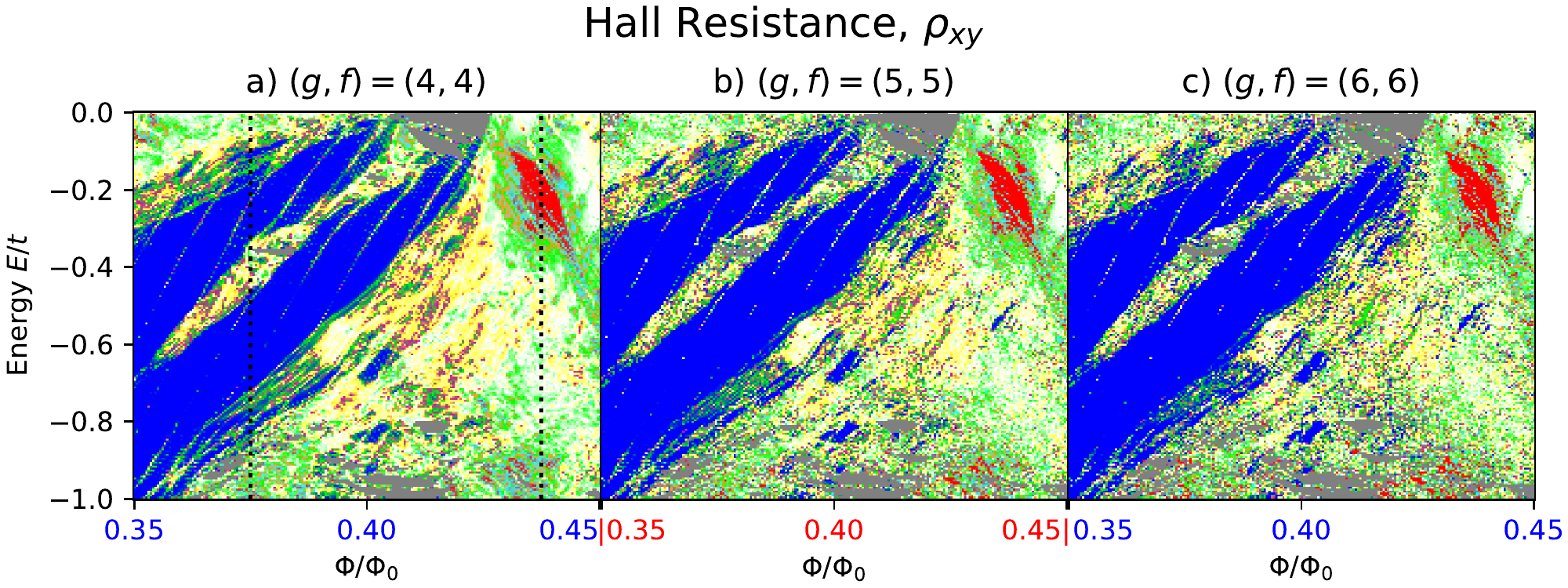}\\
        \includegraphics[width=1.0\columnwidth]{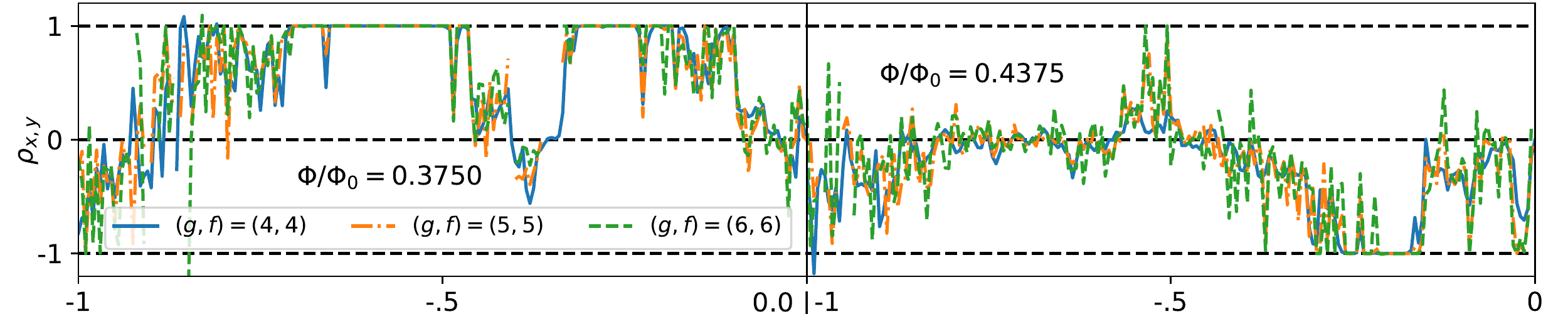}
    \par\end{centering}
\caption{Zoom in on the Hall resistance for three consecutive generations $(4,4)$, $(5,5)$ and $(6,6)$.
  The large scale structure of the edge modes is already present in generation $(4,4)$.
  The difference between the various panels is the amount of high frequency variations as seen in the two cuts at fixed $\Phi/\Phi_0$ in the lower panel.
  The color scheme is the same as in Fig.~\ref{fig:Hall_Butterfly}.
  \label{fig:Hall_Butterfly_Zoom_in}}
\end{figure}

{\it Hall Resistance:} We start with sequence (i), see Fig.~\ref{fig:Hall_Butterfly},
computing the Hall resistance for three systems of the same fractal generation $g=4$ and ascending fractal depth $(g,f)=\left(4,2\right)$, $\left(4,3\right)$,
$\left(4,4\right)$. The case of $\left(4,2\right)$ shares many features with the usual Hofstadter model,
including gaps with higher Hall conductivity, corresponding to four or five edge modes.
To quantify the edge mode counting we present a histogram of $1/|\rho_{xy}|$ in Fig.~\ref{fig:Hall_Butterfly} (d).
We note that increasing the fractal depth gaps edge modes.
The intuition is that new 'holes' introduce states inside the structure that hybridize with the actual edge states and gap them.
For the full depth fractal, we find that maximally one mode remains.
To see whether the butterfly $\left(4,4\right)$ in Fig.~\ref{fig:Hall_Butterfly} is a faithful representation of the thermodynamic limit,
we zoom in on a smaller region $0.35<\Phi/\Phi_{0}<0.45$,
$-1<E<0$ and image it with the same energy and magnetic field resolution for three consecutive system sizes $\left(4,4\right)$, $\left(5,5\right)$ and $\left(6,6\right)$,
see Fig.~\ref{fig:Hall_Butterfly_Zoom_in}.
The main finding is that all the prominent features remain intact.
Specifically, we see that the part of parameter space that hosted stable edge modes is basically unaltered in all the three plots.
We thus conclude that the maximum number of stable edge modes in a thermodynamic $(g,g)$ system appears to be 1.
Furthermore we note that the $(g,g-1)$ and $(g,-2)$ systems hosted more than one edge mode,
leading us to conjecture that at least one edge mode will always be present, irrespective of UV regularization scheme.

\begin{figure}[t]
  \begin{centering}
    \includegraphics[width=1.00\columnwidth]{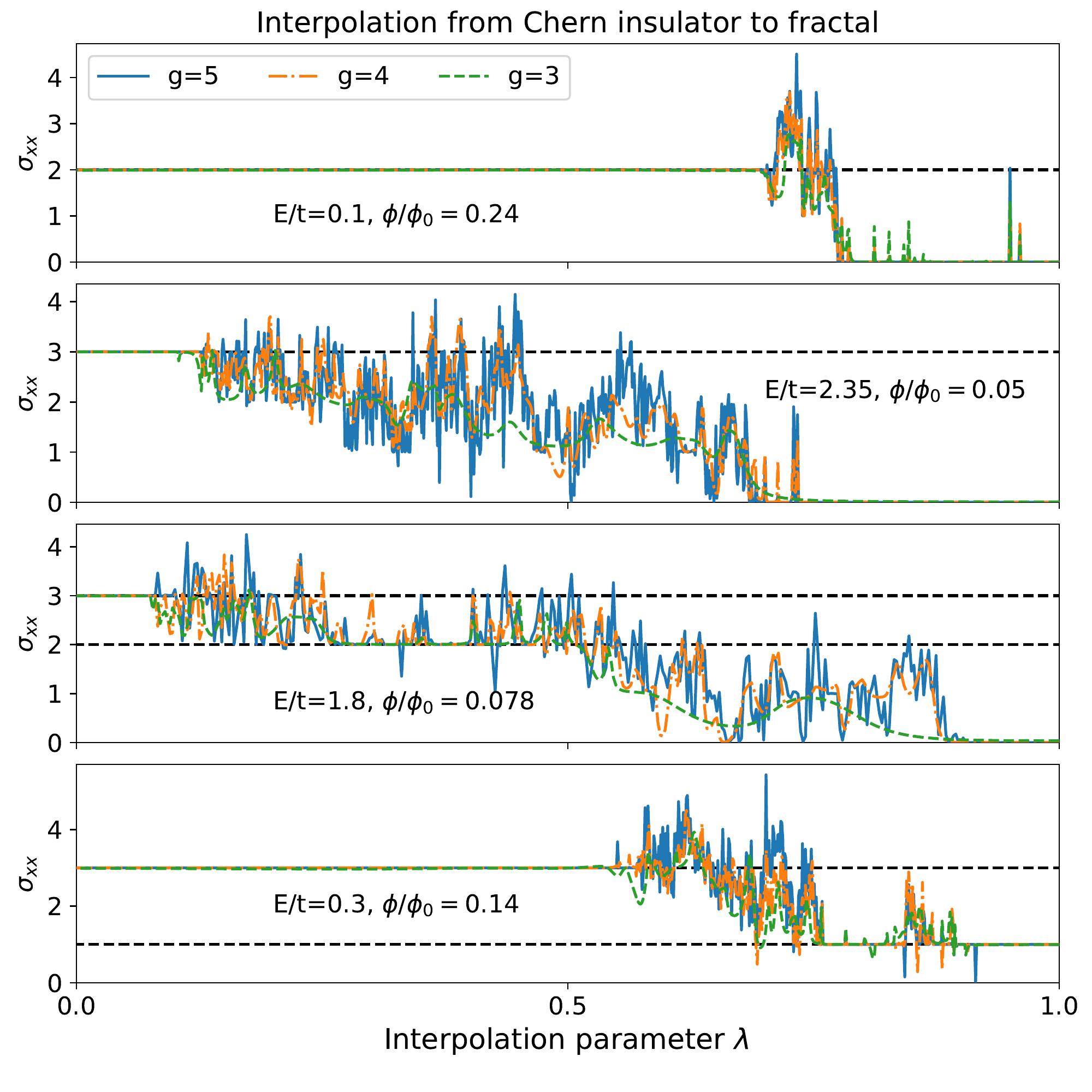}
    \par\end{centering}
    \caption{Longitudinal conductance $\sigma_{xx}=I/{V_{xx}}$ when interpolating between Chern insulator and full depth fractal.
      \label{fig:Transport_Interpolation}}
\end{figure}

\begin{figure}
  \begin{centering}
    \includegraphics[width=1.00\columnwidth]{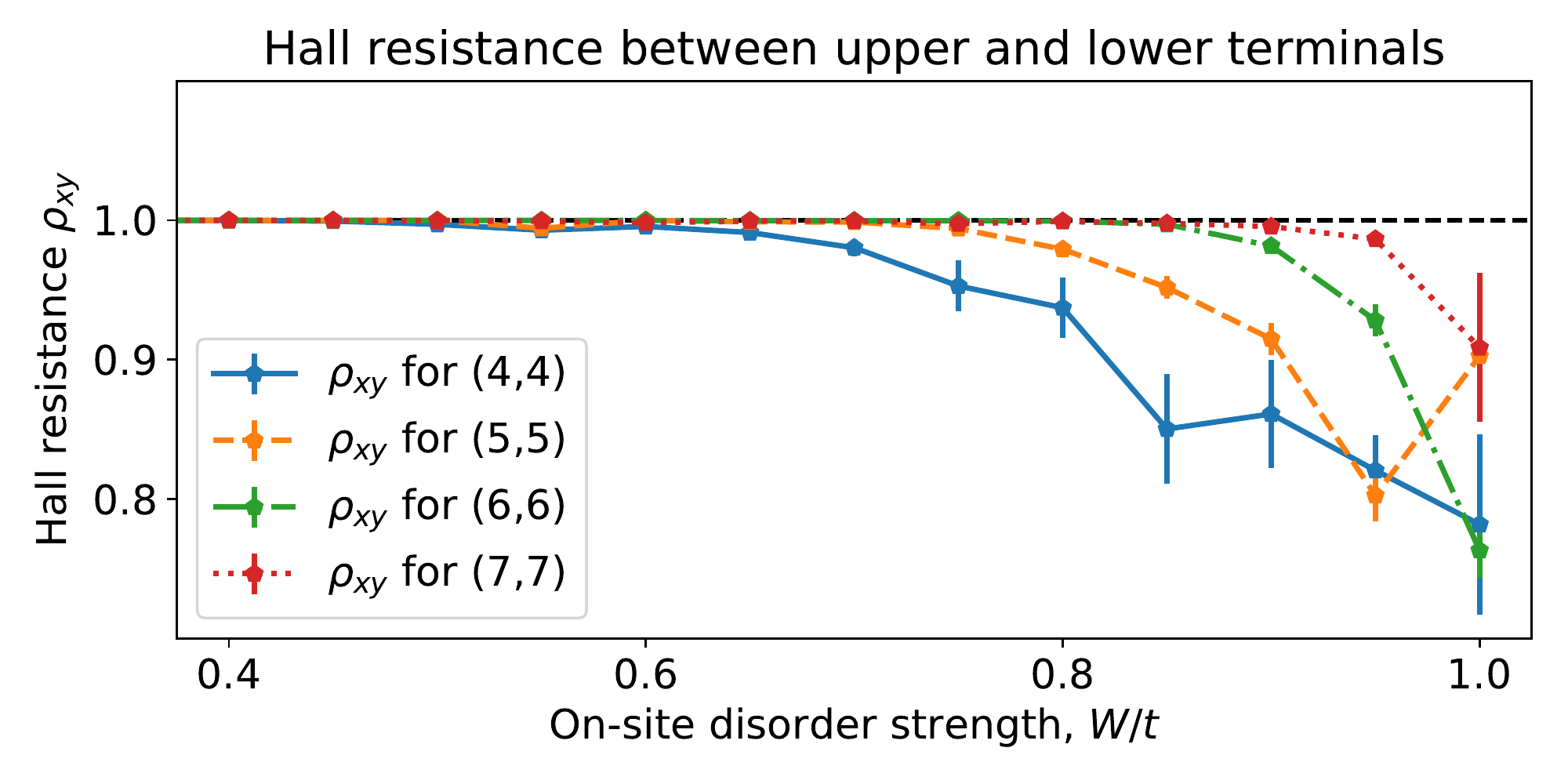}  
\par\end{centering}
\caption{Hall resistance of disordered SC fractals.
  Increasing the fractal size and depth makes the edge mode more stable with respect to the disorder strength $W$.
  The error bars show the error of the mean $\rho_{xy}$ for 20 disorder realizations.
  \label{fig:Transport_Disorder}}
\end{figure}

\begin{figure*}[th]
  \begin{centering}
    \includegraphics[width=1.0\linewidth]{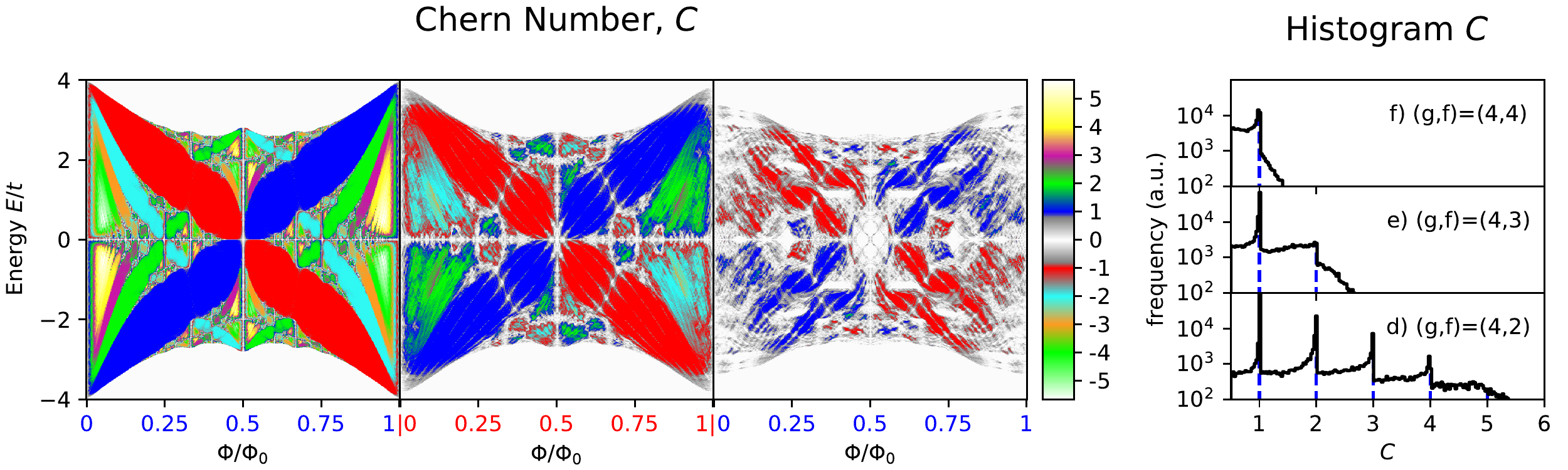} %%%Two column plot
    \par\end{centering}
    \caption{
      Direct calculation of the Chern number.
      We observe that the Chern number for $(4,2)$ and $(4,3)$ is in good agreement with $\rho_{xy}$, see Fig.~\ref{fig:Hall_Butterfly}.
      For $(4,4)$ we see deviations from the Hall resistivity which is mostly rooted in less well defined bands.
Here we use a base of 200 sites to accurately capture higher Chern numbers.
      \label{fig:Chern number}}
\end{figure*}

{\it 'Smoothly' approaching the fractal:} We have established that upon increasing fractal depth,
edge modes are annihilated with a minimum number of $1$ remaining.
To monitor the annihilation of edge modes between fractals of increasing depth we a scheme that interpolates from a shallow fractal $\mathcal{F}^1=\left(g,f_{1}\right)$ to a deeper fractal
$\mathcal{F}^2=\left(g,f_{2}\right)$, $f_{2}>f_{1}$.
For this purpose, we split the Hamiltonian into two pieces
\[H_{\mathcal{F}^1\to\mathcal{F}^2}=(1-\lambda)H_{\mathcal{F}^1}+\lambda H_{\mathcal{F}^2}=
H_{\mathcal{F}^1}-\lambda H_{\mathcal{F}^1\backslash\mathcal{F}^2}.\]
The term $H_{\mathcal{F}^1\backslash\mathcal{F}^2}$ contains all the hopping terms that are present in the shallower Hamiltonian $H_{\mathcal{F}^1}$, but not in the deeper $H_{\mathcal{F}^2}$ and $0\leq \lambda\leq1$ tunes between the two.
We consider the two-terminal transmission through the fractal for fixed $\Phi$ and study the breakdown of edge modes as a function of $\lambda$.
In all the plots we interpolate between the Hofstadter model and the full-depth fractal, see Fig.~\ref{fig:Transport_Interpolation}.

We consider the situation where the number of edge modes changes as $2\to0$ ($\Phi/\Phi_{0}=0.24$, $E=0.1$), $3\to0$ ($\Phi/\Phi_{0}=0.05$, $E=2.35$ and $\Phi/\Phi_{0}=0.078$, $E=1.8$),
and $3\to1$ ($\Phi/\Phi_{0}=0.14$, $E=0.3$).

We see that the edge modes can withstand some amount of hopping depletion before the clean edge mode is lost,
but the amount of depletion depends on the parameters $E$ and $\Phi/\Phi_{0}$.
The intuition is again that new 'holes' are introduced in the structure leading to new 'edge states' that hybridize with the actual edge states and gap them out.
We also see that the transition from edge to bulk modes is not smooth but shows high frequency oscillations in the conductance.
The frequency of the oscillations do increase with system size, but the values of $\lambda$ at which the oscillations occur,
and where there is reduction of transmission, are the same irrespective of the system sizes.
This is in agreement with the intuition that the bulk modes will have a fractal dispersion~\cite{Andrade1989,Ghez1987,Domany1983}
and thus that the bulk contribution to the conductivity should mimic this fractal pattern~\cite{Chakrabarti1997,Lin2002,vanVeen2016}.

{\it Resilience to impurities:} 
We now investigate to which extend these edge modes are stable to disorder.
We consider on-site disorder $H_I=\sum_i \epsilon_i a_i^\dagger a_i$,
where $\epsilon_i$ is uniformly distributed in the range $[-W,W]$.
We choose $\Phi/\Phi_{0}=0.25$ and $E=-1.1$ were we earlier established the existence of edge modes.
We then track $\rho_{xy}$ at four different system sizes $(g,g)$, $g=4,5,6,7$, see Fig.~\ref{fig:Transport_Disorder}.
While we found that making the fractal cuts deeper
generically decreases the number of edge modes,
the remaining single edge mode regions become increasingly stable with increasing system size.
Consequently, we conclude, that the edge modes are stable with respect to disorder,
like in a conventional quantum Hall setup.

{\it Relation to Chern Numbers:} We make a full scan of the Chern numbers, see Fig.~\ref{fig:Chern number}.
Chern numbers were computed in Ref.~\onlinecite{Brzezinska2018} for a fixed magnetic field $\Phi/\Phi_{0}=0.25$. For completeness,
we here use the same method~\cite{Kitaev2006} to compute Chern numbers for the entire $\Phi/\Phi_{0}$ and $E$ parameter space.
We compare the calculated numbers with the Hall resistances that were obtained in Fig.~\ref{fig:Hall_Butterfly}.
For $\left(4,2\right)$ and $\left(4,3\right)$, the Chern numbers coincide with $\rho_{xy}$ obtained through the transport calculations.

In the case of the full depth fractal $\left(4,4\right)$, it seems that there are larger regions with non-integer Chern number than there are in the transmission plots.
When comparing the two,
one should however keep in mind that even though the Chern number calculation is well defined for any filling of the band,
it is not expected to give integers if the Fermi energy lies within one of the bands.
To be more precise, in the Hofstadter model, narrow bands are separated by well defined band gaps.
In the fractal at finite magnetic field,
many of these bands have broken up into sub bands and the gaps between these are also quite small.
Thus interpreting what is a Chern number and what is simply ``in the band'' is more complicated.

{\it Conclusions and Outlook:} In this paper we studied edge modes and associated Hall resistivity in a Hofstadter Sierpinski fractal.
Our main finding is that increasing the fractal depth destabilizes edge modes,
but that one may never completely eliminate them.

We found that in the $(g,g)$ fractals, for sufficiently large $g$, there only are single edge modes, even in the presence of strong disorder.
Moreover, in the $(g,g-1)$ fractals, instead, $1$ and $2$ modes were present.
The reason for that is that the widest path through the lattices (which occurs \eg on the edge) is precisely one site wide.
Paths with more sites (which is already the case for $(g,g-1)$ ) enable us to see multiple edge modes.
This is most easily rationalized starting from the solutions to the Landau problem in the continuum. In the Landau gauge,
the generic wave functions have the form $\psi_{k,n} (x,y) \propto e^{-\frac{1}{2}(y-y_0(k))^{2}}H_{n}(y-y_0(k))e^{\i k x}$
($y_0(k)$ denotes where the wavefunction is centered for the respective value of $k$ and $H_n(x)$ is the Hermite polynomial of $n$-th order).
In the lowest Landau level (LL) $n=0$ $H_{0}=1$
and $\psi_{k,n}$ is reduced to $\psi_{k,0}\propto e^{-\frac{1}{2}(y-y_0(k))^2}e^{\i k x}$. This has a gaussian shape in the $y$-direction and is a plane wave in the $x$-direction.
A lattice version of this wave function can thus exist with support on a one site wide row in the $y$-direction.
For higher LLs, $n>0$, the Hermite polynomial will have nodes,
and this will cause the wave functions to change sign as a function of the distance to the edge.
To mimic the nodes and accompanying sign changes on the lattice level,
one needs at least one more site than the number of nodes in order to support the wave function.
Thus, since there is no path through the $(g,g)$ fractal that \emph{always} is more than one site wide,
only single edge modes corresponding to the lowest LL are allowed.

We stress that the existence of a one-site wide path through the fractal at maximal depth may depend on how the fractal is defined with respect to the underlying lattice.
It is therefore an interesting question to which extent edge states would be supported on a ``traditional'' fractal,
where there is no microscopic lattice scale $a$, but rather the macroscopic scale $L$ is held fixed.

In relation of other fractal lattices in magnetic fields,
we speculate that the Hausdorff dimension and/or the connectivity plays a role.
We note for instance that the Sierpinski Triangle/Gasket does not seem to host stable edge modes\cite{Brzezinska2018}.
This system has a lower dimension than the carpet, but also the connectivity is finite, whereas in the carpet it is extensive.

In the future, it would be interesting to see how other fractal systems may support topologically ordered states, for instance, by considering three dimensional topological insulators on 3D Sierpinski gaskets. 

After the completion and submission of this work, Ref.~\onlinecite{Iliasov2019b} appeared which treats a similar problem using the Kubo-Bastin formula for Hall conductivity. They obtain similar results as in our paper.

{\it Acknowledgements:}
  We acknowledge the SFI/HEA Irish Centre for High-End Computing (ICHEC) for the provision of computational facilities and support through project nmphy013b.
This work is part of the D-ITP consortium,
a program of the Netherlands Organisation for Scientific Research (NWO) that is funded by the Dutch Ministry of Education, Culture and Science (OCW).

\bibliography{Edge_States_In_Fractals}

\begin{thebibliography}{36}
\expandafter\ifx\csname natexlab\endcsname\relax\def\natexlab#1{#1}\fi
\expandafter\ifx\csname bibnamefont\endcsname\relax
  \def\bibnamefont#1{#1}\fi
\expandafter\ifx\csname bibfnamefont\endcsname\relax
  \def\bibfnamefont#1{#1}\fi
\expandafter\ifx\csname citenamefont\endcsname\relax
  \def\citenamefont#1{#1}\fi
\expandafter\ifx\csname url\endcsname\relax
  \def\url#1{\texttt{#1}}\fi
\expandafter\ifx\csname urlprefix\endcsname\relax\def\urlprefix{URL }\fi
\providecommand{\bibinfo}[2]{#2}
\providecommand{\eprint}[2][]{\url{#2}}

\bibitem[{\citenamefont{Bernevig and Hughes}(2013)}]{bernevig2013topological}
\bibinfo{author}{\bibfnamefont{B.~A.} \bibnamefont{Bernevig}} \bibnamefont{and}
  \bibinfo{author}{\bibfnamefont{T.~L.} \bibnamefont{Hughes}},
  \emph{\bibinfo{title}{Topological insulators and topological
  superconductors}} (\bibinfo{publisher}{Princeton university press},
  \bibinfo{year}{2013}).

\bibitem[{\citenamefont{Ryu et~al.}(2010)\citenamefont{Ryu, Schnyder, Furusaki,
  and Ludwig}}]{ryu2010topological}
\bibinfo{author}{\bibfnamefont{S.}~\bibnamefont{Ryu}},
  \bibinfo{author}{\bibfnamefont{A.~P.} \bibnamefont{Schnyder}},
  \bibinfo{author}{\bibfnamefont{A.}~\bibnamefont{Furusaki}}, \bibnamefont{and}
  \bibinfo{author}{\bibfnamefont{A.~W.} \bibnamefont{Ludwig}},
  \bibinfo{journal}{New Journal of Physics} \textbf{\bibinfo{volume}{12}},
  \bibinfo{pages}{065010} (\bibinfo{year}{2010}).

\bibitem[{\citenamefont{Qi and Zhang}(2011)}]{qi2011topological}
\bibinfo{author}{\bibfnamefont{X.-L.} \bibnamefont{Qi}} \bibnamefont{and}
  \bibinfo{author}{\bibfnamefont{S.-C.} \bibnamefont{Zhang}},
  \bibinfo{journal}{Reviews of Modern Physics} \textbf{\bibinfo{volume}{83}},
  \bibinfo{pages}{1057} (\bibinfo{year}{2011}).

\bibitem[{\citenamefont{Kitaev}(2009)}]{kitaev2009periodic}
\bibinfo{author}{\bibfnamefont{A.}~\bibnamefont{Kitaev}}, in
  \emph{\bibinfo{booktitle}{AIP Conference Proceedings}}
  (\bibinfo{organization}{AIP}, \bibinfo{year}{2009}), vol.
  \bibinfo{volume}{1134}, pp. \bibinfo{pages}{22--30}.

\bibitem[{\citenamefont{Fu}(2011)}]{fu2011topological}
\bibinfo{author}{\bibfnamefont{L.}~\bibnamefont{Fu}},
  \bibinfo{journal}{Physical Review Letters} \textbf{\bibinfo{volume}{106}},
  \bibinfo{pages}{106802} (\bibinfo{year}{2011}).

\bibitem[{\citenamefont{Diehl et~al.}(2011)\citenamefont{Diehl, Rico, Baranov,
  and Zoller}}]{diehl2011topology}
\bibinfo{author}{\bibfnamefont{S.}~\bibnamefont{Diehl}},
  \bibinfo{author}{\bibfnamefont{E.}~\bibnamefont{Rico}},
  \bibinfo{author}{\bibfnamefont{M.~A.} \bibnamefont{Baranov}},
  \bibnamefont{and} \bibinfo{author}{\bibfnamefont{P.}~\bibnamefont{Zoller}},
  \bibinfo{journal}{Nature Physics} \textbf{\bibinfo{volume}{7}},
  \bibinfo{pages}{971} (\bibinfo{year}{2011}).

\bibitem[{\citenamefont{Lindner et~al.}(2011)\citenamefont{Lindner, Refael, and
  Galitski}}]{lindner2011floquet}
\bibinfo{author}{\bibfnamefont{N.~H.} \bibnamefont{Lindner}},
  \bibinfo{author}{\bibfnamefont{G.}~\bibnamefont{Refael}}, \bibnamefont{and}
  \bibinfo{author}{\bibfnamefont{V.}~\bibnamefont{Galitski}},
  \bibinfo{journal}{Nature Physics} \textbf{\bibinfo{volume}{7}},
  \bibinfo{pages}{490} (\bibinfo{year}{2011}).

\bibitem[{\citenamefont{Benalcazar et~al.}(2017)\citenamefont{Benalcazar,
  Bernevig, and Hughes}}]{benalcazar2017quantized}
\bibinfo{author}{\bibfnamefont{W.~A.} \bibnamefont{Benalcazar}},
  \bibinfo{author}{\bibfnamefont{B.~A.} \bibnamefont{Bernevig}},
  \bibnamefont{and} \bibinfo{author}{\bibfnamefont{T.~L.}
  \bibnamefont{Hughes}}, \bibinfo{journal}{Science}
  \textbf{\bibinfo{volume}{357}}, \bibinfo{pages}{61} (\bibinfo{year}{2017}).

\bibitem[{\citenamefont{Schindler et~al.}(2018)\citenamefont{Schindler, Cook,
  Vergniory, Wang, Parkin, Bernevig, and Neupert}}]{schindler2018higher}
\bibinfo{author}{\bibfnamefont{F.}~\bibnamefont{Schindler}},
  \bibinfo{author}{\bibfnamefont{A.~M.} \bibnamefont{Cook}},
  \bibinfo{author}{\bibfnamefont{M.~G.} \bibnamefont{Vergniory}},
  \bibinfo{author}{\bibfnamefont{Z.}~\bibnamefont{Wang}},
  \bibinfo{author}{\bibfnamefont{S.~S.} \bibnamefont{Parkin}},
  \bibinfo{author}{\bibfnamefont{B.~A.} \bibnamefont{Bernevig}},
  \bibnamefont{and} \bibinfo{author}{\bibfnamefont{T.}~\bibnamefont{Neupert}},
  \bibinfo{journal}{Science advances} \textbf{\bibinfo{volume}{4}},
  \bibinfo{pages}{eaat0346} (\bibinfo{year}{2018}).

\bibitem[{\citenamefont{McMullen}(1984)}]{mcmullen1984hausdorff}
\bibinfo{author}{\bibfnamefont{C.}~\bibnamefont{McMullen}},
  \bibinfo{journal}{Nagoya Mathematical Journal} \textbf{\bibinfo{volume}{96}},
  \bibinfo{pages}{1} (\bibinfo{year}{1984}).

\bibitem[{\citenamefont{Shang et~al.}(2015)\citenamefont{Shang, Wang, Chen,
  Dai, Zhou, Kuttner, Hilt, Shao, Gottfried, and Wu}}]{Shang2015}
\bibinfo{author}{\bibfnamefont{J.}~\bibnamefont{Shang}},
  \bibinfo{author}{\bibfnamefont{Y.}~\bibnamefont{Wang}},
  \bibinfo{author}{\bibfnamefont{M.}~\bibnamefont{Chen}},
  \bibinfo{author}{\bibfnamefont{J.}~\bibnamefont{Dai}},
  \bibinfo{author}{\bibfnamefont{X.}~\bibnamefont{Zhou}},
  \bibinfo{author}{\bibfnamefont{J.}~\bibnamefont{Kuttner}},
  \bibinfo{author}{\bibfnamefont{G.}~\bibnamefont{Hilt}},
  \bibinfo{author}{\bibfnamefont{X.}~\bibnamefont{Shao}},
  \bibinfo{author}{\bibfnamefont{J.~M.} \bibnamefont{Gottfried}},
  \bibnamefont{and} \bibinfo{author}{\bibfnamefont{K.}~\bibnamefont{Wu}},
  \bibinfo{journal}{Nature chemistry} \textbf{\bibinfo{volume}{7}},
  \bibinfo{pages}{389} (\bibinfo{year}{2015}).

\bibitem[{\citenamefont{Tait}(2015)}]{Tait2015}
\bibinfo{author}{\bibfnamefont{S.~L.} \bibnamefont{Tait}},
  \bibinfo{journal}{Nature chemistry} \textbf{\bibinfo{volume}{7}},
  \bibinfo{pages}{370} (\bibinfo{year}{2015}).

\bibitem[{\citenamefont{Nieckarz and Szabelski}(2016)}]{Nieckarz2016}
\bibinfo{author}{\bibfnamefont{D.}~\bibnamefont{Nieckarz}} \bibnamefont{and}
  \bibinfo{author}{\bibfnamefont{P.}~\bibnamefont{Szabelski}},
  \bibinfo{journal}{Chemical Communications} \textbf{\bibinfo{volume}{52}},
  \bibinfo{pages}{11642} (\bibinfo{year}{2016}).

\bibitem[{\citenamefont{Jiang et~al.}(2017)\citenamefont{Jiang, Li, Wang, Liu,
  Yuan, Chen, Wang, Newkome, Sun, Li et~al.}}]{Jiang2017}
\bibinfo{author}{\bibfnamefont{Z.}~\bibnamefont{Jiang}},
  \bibinfo{author}{\bibfnamefont{Y.}~\bibnamefont{Li}},
  \bibinfo{author}{\bibfnamefont{M.}~\bibnamefont{Wang}},
  \bibinfo{author}{\bibfnamefont{D.}~\bibnamefont{Liu}},
  \bibinfo{author}{\bibfnamefont{J.}~\bibnamefont{Yuan}},
  \bibinfo{author}{\bibfnamefont{M.}~\bibnamefont{Chen}},
  \bibinfo{author}{\bibfnamefont{J.}~\bibnamefont{Wang}},
  \bibinfo{author}{\bibfnamefont{G.~R.} \bibnamefont{Newkome}},
  \bibinfo{author}{\bibfnamefont{W.}~\bibnamefont{Sun}},
  \bibinfo{author}{\bibfnamefont{X.}~\bibnamefont{Li}}, \bibnamefont{et~al.},
  \bibinfo{journal}{Angewandte Chemie} \textbf{\bibinfo{volume}{129}},
  \bibinfo{pages}{11608} (\bibinfo{year}{2017}).

\bibitem[{\citenamefont{Sun et~al.}(2015)\citenamefont{Sun, Cai, Ma, Yuan, and
  Xu}}]{Sun2015}
\bibinfo{author}{\bibfnamefont{Q.}~\bibnamefont{Sun}},
  \bibinfo{author}{\bibfnamefont{L.}~\bibnamefont{Cai}},
  \bibinfo{author}{\bibfnamefont{H.}~\bibnamefont{Ma}},
  \bibinfo{author}{\bibfnamefont{C.}~\bibnamefont{Yuan}}, \bibnamefont{and}
  \bibinfo{author}{\bibfnamefont{W.}~\bibnamefont{Xu}},
  \bibinfo{journal}{Chemical Communications} \textbf{\bibinfo{volume}{51}},
  \bibinfo{pages}{14164} (\bibinfo{year}{2015}).

\bibitem[{\citenamefont{Zhang et~al.}(2015)\citenamefont{Zhang, Li, Gu, Wang,
  Nieckarz, Szabelski, He, Wang, Xie, Shen et~al.}}]{Zhang2015}
\bibinfo{author}{\bibfnamefont{X.}~\bibnamefont{Zhang}},
  \bibinfo{author}{\bibfnamefont{N.}~\bibnamefont{Li}},
  \bibinfo{author}{\bibfnamefont{G.-C.} \bibnamefont{Gu}},
  \bibinfo{author}{\bibfnamefont{H.}~\bibnamefont{Wang}},
  \bibinfo{author}{\bibfnamefont{D.}~\bibnamefont{Nieckarz}},
  \bibinfo{author}{\bibfnamefont{P.}~\bibnamefont{Szabelski}},
  \bibinfo{author}{\bibfnamefont{Y.}~\bibnamefont{He}},
  \bibinfo{author}{\bibfnamefont{Y.}~\bibnamefont{Wang}},
  \bibinfo{author}{\bibfnamefont{C.}~\bibnamefont{Xie}},
  \bibinfo{author}{\bibfnamefont{Z.-Y.} \bibnamefont{Shen}},
  \bibnamefont{et~al.}, \bibinfo{journal}{ACS nano}
  \textbf{\bibinfo{volume}{9}}, \bibinfo{pages}{11909} (\bibinfo{year}{2015}).

\bibitem[{\citenamefont{Zhang et~al.}(2016)\citenamefont{Zhang, Li, Liu, Gu,
  Li, Tang, Peng, Hou, and Wang}}]{Zhang2016}
\bibinfo{author}{\bibfnamefont{X.}~\bibnamefont{Zhang}},
  \bibinfo{author}{\bibfnamefont{N.}~\bibnamefont{Li}},
  \bibinfo{author}{\bibfnamefont{L.}~\bibnamefont{Liu}},
  \bibinfo{author}{\bibfnamefont{G.}~\bibnamefont{Gu}},
  \bibinfo{author}{\bibfnamefont{C.}~\bibnamefont{Li}},
  \bibinfo{author}{\bibfnamefont{H.}~\bibnamefont{Tang}},
  \bibinfo{author}{\bibfnamefont{L.}~\bibnamefont{Peng}},
  \bibinfo{author}{\bibfnamefont{S.}~\bibnamefont{Hou}}, \bibnamefont{and}
  \bibinfo{author}{\bibfnamefont{Y.}~\bibnamefont{Wang}},
  \bibinfo{journal}{Chemical Communications} \textbf{\bibinfo{volume}{52}},
  \bibinfo{pages}{10578} (\bibinfo{year}{2016}).

\bibitem[{\citenamefont{Li et~al.}(2017{\natexlab{a}})\citenamefont{Li, Zhang,
  Li, Wang, Yang, Gu, Zhang, Hou, Peng, Wu et~al.}}]{Li2017a}
\bibinfo{author}{\bibfnamefont{C.}~\bibnamefont{Li}},
  \bibinfo{author}{\bibfnamefont{X.}~\bibnamefont{Zhang}},
  \bibinfo{author}{\bibfnamefont{N.}~\bibnamefont{Li}},
  \bibinfo{author}{\bibfnamefont{Y.}~\bibnamefont{Wang}},
  \bibinfo{author}{\bibfnamefont{J.}~\bibnamefont{Yang}},
  \bibinfo{author}{\bibfnamefont{G.}~\bibnamefont{Gu}},
  \bibinfo{author}{\bibfnamefont{Y.}~\bibnamefont{Zhang}},
  \bibinfo{author}{\bibfnamefont{S.}~\bibnamefont{Hou}},
  \bibinfo{author}{\bibfnamefont{L.}~\bibnamefont{Peng}},
  \bibinfo{author}{\bibfnamefont{K.}~\bibnamefont{Wu}}, \bibnamefont{et~al.},
  \bibinfo{journal}{Journal of the American Chemical Society}
  \textbf{\bibinfo{volume}{139}}, \bibinfo{pages}{13749}
  (\bibinfo{year}{2017}{\natexlab{a}}).

\bibitem[{\citenamefont{Li et~al.}(2017{\natexlab{b}})\citenamefont{Li, Gu,
  Zhang, Song, Zhang, Teo, Peng, Hou, and Wang}}]{Li2017b}
\bibinfo{author}{\bibfnamefont{N.}~\bibnamefont{Li}},
  \bibinfo{author}{\bibfnamefont{G.}~\bibnamefont{Gu}},
  \bibinfo{author}{\bibfnamefont{X.}~\bibnamefont{Zhang}},
  \bibinfo{author}{\bibfnamefont{D.}~\bibnamefont{Song}},
  \bibinfo{author}{\bibfnamefont{Y.}~\bibnamefont{Zhang}},
  \bibinfo{author}{\bibfnamefont{B.~K.} \bibnamefont{Teo}},
  \bibinfo{author}{\bibfnamefont{L.-m.} \bibnamefont{Peng}},
  \bibinfo{author}{\bibfnamefont{S.}~\bibnamefont{Hou}}, \bibnamefont{and}
  \bibinfo{author}{\bibfnamefont{Y.}~\bibnamefont{Wang}},
  \bibinfo{journal}{Chemical Communications} \textbf{\bibinfo{volume}{53}},
  \bibinfo{pages}{3469} (\bibinfo{year}{2017}{\natexlab{b}}).

\bibitem[{\citenamefont{Kempkes et~al.}(2019)\citenamefont{Kempkes, Slot,
  Freeney, Zevenhuizen, Vanmaekelbergh, Swart, and Smith}}]{Kempkes2019}
\bibinfo{author}{\bibfnamefont{S.}~\bibnamefont{Kempkes}},
  \bibinfo{author}{\bibfnamefont{M.}~\bibnamefont{Slot}},
  \bibinfo{author}{\bibfnamefont{S.}~\bibnamefont{Freeney}},
  \bibinfo{author}{\bibfnamefont{S.}~\bibnamefont{Zevenhuizen}},
  \bibinfo{author}{\bibfnamefont{D.}~\bibnamefont{Vanmaekelbergh}},
  \bibinfo{author}{\bibfnamefont{I.}~\bibnamefont{Swart}}, \bibnamefont{and}
  \bibinfo{author}{\bibfnamefont{C.~M.} \bibnamefont{Smith}},
  \bibinfo{journal}{Nature Physics} \textbf{\bibinfo{volume}{15}},
  \bibinfo{pages}{127} (\bibinfo{year}{2019}).

\bibitem[{\citenamefont{van Veen et~al.}(2016)\citenamefont{van Veen, Yuan,
  Katsnelson, Polini, and Tomadin}}]{vanVeen2016}
\bibinfo{author}{\bibfnamefont{E.}~\bibnamefont{van Veen}},
  \bibinfo{author}{\bibfnamefont{S.}~\bibnamefont{Yuan}},
  \bibinfo{author}{\bibfnamefont{M.~I.} \bibnamefont{Katsnelson}},
  \bibinfo{author}{\bibfnamefont{M.}~\bibnamefont{Polini}}, \bibnamefont{and}
  \bibinfo{author}{\bibfnamefont{A.}~\bibnamefont{Tomadin}},
  \bibinfo{journal}{Physical Review B} \textbf{\bibinfo{volume}{93}},
  \bibinfo{pages}{115428} (\bibinfo{year}{2016}).

\bibitem[{\citenamefont{van Veen et~al.}(2017)\citenamefont{van Veen, Tomadin,
  Polini, Katsnelson, and Yuan}}]{vanVeen2017}
\bibinfo{author}{\bibfnamefont{E.}~\bibnamefont{van Veen}},
  \bibinfo{author}{\bibfnamefont{A.}~\bibnamefont{Tomadin}},
  \bibinfo{author}{\bibfnamefont{M.}~\bibnamefont{Polini}},
  \bibinfo{author}{\bibfnamefont{M.~I.} \bibnamefont{Katsnelson}},
  \bibnamefont{and} \bibinfo{author}{\bibfnamefont{S.}~\bibnamefont{Yuan}},
  \bibinfo{journal}{Physical Review B} \textbf{\bibinfo{volume}{96}},
  \bibinfo{pages}{235438} (\bibinfo{year}{2017}).

\bibitem[{\citenamefont{Westerhout et~al.}(2018)\citenamefont{Westerhout, van
  Veen, Katsnelson, and Yuan}}]{Westerhout2018}
\bibinfo{author}{\bibfnamefont{T.}~\bibnamefont{Westerhout}},
  \bibinfo{author}{\bibfnamefont{E.}~\bibnamefont{van Veen}},
  \bibinfo{author}{\bibfnamefont{M.~I.} \bibnamefont{Katsnelson}},
  \bibnamefont{and} \bibinfo{author}{\bibfnamefont{S.}~\bibnamefont{Yuan}},
  \bibinfo{journal}{Physical Review B} \textbf{\bibinfo{volume}{97}},
  \bibinfo{pages}{205434} (\bibinfo{year}{2018}).

\bibitem[{\citenamefont{Sticlet and Akhmerov}(2016)}]{Sticlet2016}
\bibinfo{author}{\bibfnamefont{D.}~\bibnamefont{Sticlet}} \bibnamefont{and}
  \bibinfo{author}{\bibfnamefont{A.}~\bibnamefont{Akhmerov}},
  \bibinfo{journal}{Physical Review B} \textbf{\bibinfo{volume}{94}},
  \bibinfo{pages}{161115(R)} (\bibinfo{year}{2016}).

\bibitem[{\citenamefont{Iliasov
  et~al.}(2019{\natexlab{a}})\citenamefont{Iliasov, Katsnelson, and
  Yuan}}]{Iliasov2019}
\bibinfo{author}{\bibfnamefont{A.~A.} \bibnamefont{Iliasov}},
  \bibinfo{author}{\bibfnamefont{M.~I.} \bibnamefont{Katsnelson}},
  \bibnamefont{and} \bibinfo{author}{\bibfnamefont{S.}~\bibnamefont{Yuan}},
  \bibinfo{journal}{Physical Review B} \textbf{\bibinfo{volume}{99}},
  \bibinfo{pages}{075402} (\bibinfo{year}{2019}{\natexlab{a}}).

\bibitem[{\citenamefont{Aoki}(1990)}]{Aoki1990}
\bibinfo{author}{\bibfnamefont{H.}~\bibnamefont{Aoki}}, \bibinfo{journal}{Phys.
  Rev. B} \textbf{\bibinfo{volume}{42}}, \bibinfo{pages}{6869}
  (\bibinfo{year}{1990}),
  \urlprefix\url{https://link.aps.org/doi/10.1103/PhysRevB.42.6869}.

\bibitem[{\citenamefont{Aoki}(1992)}]{Aoki1992}
\bibinfo{author}{\bibfnamefont{H.}~\bibnamefont{Aoki}},
  \bibinfo{journal}{Surface science} \textbf{\bibinfo{volume}{263}},
  \bibinfo{pages}{137} (\bibinfo{year}{1992}).

\bibitem[{\citenamefont{Brzezi{\'n}ska
  et~al.}(2018)\citenamefont{Brzezi{\'n}ska, Cook, and
  Neupert}}]{Brzezinska2018}
\bibinfo{author}{\bibfnamefont{M.}~\bibnamefont{Brzezi{\'n}ska}},
  \bibinfo{author}{\bibfnamefont{A.~M.} \bibnamefont{Cook}}, \bibnamefont{and}
  \bibinfo{author}{\bibfnamefont{T.}~\bibnamefont{Neupert}},
  \bibinfo{journal}{Physical Review B} \textbf{\bibinfo{volume}{98}},
  \bibinfo{pages}{205116} (\bibinfo{year}{2018}).

\bibitem[{\citenamefont{Datta}(1997)}]{datta1997electronic}
\bibinfo{author}{\bibfnamefont{S.}~\bibnamefont{Datta}},
  \emph{\bibinfo{title}{Electronic transport in mesoscopic systems}}
  (\bibinfo{publisher}{Cambridge university press}, \bibinfo{year}{1997}).

\bibitem[{\citenamefont{Andrade and Schellnhuber}(1989)}]{Andrade1989}
\bibinfo{author}{\bibfnamefont{R.}~\bibnamefont{Andrade}} \bibnamefont{and}
  \bibinfo{author}{\bibfnamefont{H.}~\bibnamefont{Schellnhuber}},
  \bibinfo{journal}{EPL (Europhysics Letters)} \textbf{\bibinfo{volume}{10}},
  \bibinfo{pages}{73} (\bibinfo{year}{1989}).

\bibitem[{\citenamefont{Ghez et~al.}(1987)\citenamefont{Ghez, Wang, Rammal,
  Pannetier, and Bellissard}}]{Ghez1987}
\bibinfo{author}{\bibfnamefont{J.~M.} \bibnamefont{Ghez}},
  \bibinfo{author}{\bibfnamefont{Y.~Y.} \bibnamefont{Wang}},
  \bibinfo{author}{\bibfnamefont{R.}~\bibnamefont{Rammal}},
  \bibinfo{author}{\bibfnamefont{B.}~\bibnamefont{Pannetier}},
  \bibnamefont{and}
  \bibinfo{author}{\bibfnamefont{J.}~\bibnamefont{Bellissard}},
  \bibinfo{journal}{Solid state communications} \textbf{\bibinfo{volume}{64}},
  \bibinfo{pages}{1291} (\bibinfo{year}{1987}).

\bibitem[{\citenamefont{Domany et~al.}(1983)\citenamefont{Domany, Alexander,
  Bensimon, and Kadanoff}}]{Domany1983}
\bibinfo{author}{\bibfnamefont{E.}~\bibnamefont{Domany}},
  \bibinfo{author}{\bibfnamefont{S.}~\bibnamefont{Alexander}},
  \bibinfo{author}{\bibfnamefont{D.}~\bibnamefont{Bensimon}}, \bibnamefont{and}
  \bibinfo{author}{\bibfnamefont{L.~P.} \bibnamefont{Kadanoff}},
  \bibinfo{journal}{Physical Review B} \textbf{\bibinfo{volume}{28}},
  \bibinfo{pages}{3110} (\bibinfo{year}{1983}).

\bibitem[{\citenamefont{Chakrabarti and Bhattacharyya}(1997)}]{Chakrabarti1997}
\bibinfo{author}{\bibfnamefont{A.}~\bibnamefont{Chakrabarti}} \bibnamefont{and}
  \bibinfo{author}{\bibfnamefont{B.}~\bibnamefont{Bhattacharyya}},
  \bibinfo{journal}{Physical Review B} \textbf{\bibinfo{volume}{56}},
  \bibinfo{pages}{13768} (\bibinfo{year}{1997}).

\bibitem[{\citenamefont{Lin et~al.}(2002)\citenamefont{Lin, Cao, Liu, and
  Hui}}]{Lin2002}
\bibinfo{author}{\bibfnamefont{Z.}~\bibnamefont{Lin}},
  \bibinfo{author}{\bibfnamefont{Y.}~\bibnamefont{Cao}},
  \bibinfo{author}{\bibfnamefont{Y.}~\bibnamefont{Liu}}, \bibnamefont{and}
  \bibinfo{author}{\bibfnamefont{P.~M.} \bibnamefont{Hui}},
  \bibinfo{journal}{Physical Review B} \textbf{\bibinfo{volume}{66}},
  \bibinfo{pages}{045311} (\bibinfo{year}{2002}).

\bibitem[{\citenamefont{Kitaev}(2006)}]{Kitaev2006}
\bibinfo{author}{\bibfnamefont{A.}~\bibnamefont{Kitaev}},
  \bibinfo{journal}{Annals of Physics} \textbf{\bibinfo{volume}{321}},
  \bibinfo{pages}{2} (\bibinfo{year}{2006}).

\bibitem[{\citenamefont{Iliasov
  et~al.}(2019{\natexlab{b}})\citenamefont{Iliasov, Katsnelson, and
  Yuan}}]{Iliasov2019b}
\bibinfo{author}{\bibfnamefont{A.~A.} \bibnamefont{Iliasov}},
  \bibinfo{author}{\bibfnamefont{M.~I.} \bibnamefont{Katsnelson}},
  \bibnamefont{and} \bibinfo{author}{\bibfnamefont{S.}~\bibnamefont{Yuan}},
  \bibinfo{journal}{arXiv preprint arXiv:1907.09310}
  (\bibinfo{year}{2019}{\natexlab{b}}).

\end{thebibliography}
\bibliographystyle{apsrev}

\end{document}